\documentclass[a4paper,11pt]{article}
\usepackage{pos}

\title{Probing the multi-scale dynamical interaction between heavy quarks and the QGP using JETSCAPE}

\author*[a]{Wenkai Fan}
\author[b]{Gojko Vujanovic for the JETSCAPE Collaboration}

\affiliation[a]{Department of Physics, Duke University\\Durham, NC 27708, USA}
\affiliation[b]{Department of Physics and Astronomy, Wayne State University\\ Detroit, Michigan 48201, USA}

\emailAdd{wf39@duke.edu}

\abstract{The dynamics of shower development for a jet traveling through the QGP involves a variety of scales, one of them being the heavy quark mass. Even though the mass of the heavy quarks plays a subdominant role during the high virtuality portion of the jet evolution, it does affect longitudinal drag and diffusion, stimulating additional radiation from heavy quarks \cite{abir2016}. These emissions partially compensate the reduction in radiation from the dead cone effect. In the lower virtuality part of the shower, when the mass is comparable to the transverse momenta of the partons, scattering and radiation processes off heavy quarks differ from those off light quarks. All these factors result in a different nuclear modification factor for heavy versus light flavors and thus for heavy-flavor tagged jets.

In this study, the heavy quark shower evolution and the fluid dynamical medium are modeled on an event by event basis using the JETSCAPE Framework \cite{putschke2019}. We present a multi-stage calculation that explores the  differences between various heavy quark energy-loss mechanisms within a realistically expanding quark-gluon plasma (QGP). Inside the QGP, the highly virtual and energetic portion of the shower is modeled using the MATTER generator, while the LBT generator models the showers induced by energetic and close-to-on-shell heavy quarks. Energy-momentum exchange with the medium, essential for the study of jet modification, proceeds using a weak coupling recoil approach. The JETSCAPE framework allows for transitions, on the level of individual partons, from one energy-loss prescription to the other depending on the parton’s energy and virtuality and the local density. This allows us to explore the effect and interplay between the different regimes of energy loss on the propagation and radiation from hard heavy quarks in a dense medium.}

\FullConference{%
  HardProbes2020\\
  1-6 June 2020\\
  Austin, Texas}

\begin{document}
\maketitle

\section{Introduction}

JETSCAPE (Jet Energy-loss Tomography with a Statistically and Computationally Advanced Program
Envelope) is a modular, flexible, publicly available event-generator framework modeling all aspects of heavy
ion collisions. In this study we focus on the description of the interaction between heavy quarks and the QGP. The interaction is modeled via a two-step approach describing both light and heavy flavor partons: the high virtuality (and high energy) portion of parton interaction with the QGP is described using the higher twist formalism \cite{wang2001, majumder2012} implemented in MATTER  (Modular All Twist Transverse-scattering Elastic-drag and Radiation) \cite{majumder2013, abir2016}, while the low virtuality (and high energy) quenching of partons is described via Linear Boltzmann Transport (LBT) \cite{luo2018}. A complete study of high-$p_T$ light
flavor hadronic observables within the JETSCAPE framework can be found in Ref.\cite{amit2019} which is used in turn to fix all the parameters for heavy-flavors’ interaction with the QGP explored within the present contribution, without any additional tuning.

\section{Simulation setup}
The event-by-event hydrodynamical simulations used throughout this study are calibrated using an established Bayesian model-to-data comparison \cite{bernhand2019}. That calibration uses the TRENTo initial conditions, followed by free-streaming and (2+1)-D viscous hydrodynamics. We use the parameters corresponding to the maximum likelihood of the posterior distribution to generate the the QGP medium for Pb-Pb collisions at $\sqrt{s_{NN}}=5.02$ TeV in the 0-10\% centrality class. Hard partons, initially produced via PYTHIA, are then allowed to exchange their energy and momentum with the pre-simulated dynamically evolving QGP, using either MATTER or LBT as the quenching formalism, depending on the virtuality of each parton. Partons with virtuality $Q > Q_0$ (with $Q_0$ being the switching scale) are handled by MATTER, while all partons with $Q < Q_0$ are given to LBT. The switching virtuality $Q_0$ found to best describe light flavor observables was $Q_0 = 2$~GeV. This value of $Q_0$ was also used for heavy flavor evolution within the QGP.

Based on the higher twist formalism, MATTER is a virtuality-ordered Monte Carlo (MC) event generator describing parton splittings according to a generalized Sudakov form factor, which includes vacuum and in-medium contributions. In this calculation, the in-medium contribution to the Sudakov form factor accounts for transverse momentum broadening of partons ($\hat{q}$) as they travel through the QGP. An effective strong coupling $\alpha_s = 0.25$ is used to determine $\hat{q}$, which was tuned using light flavor observables. 

The LBT portion of heavy flavor interaction with the QGP relies on solving the linearized Boltzmann equation, containing 2 → 2 and 2 → 3 processes. The 2 → 2 processes consist of leading order perturbative QCD scatterings between hard and thermal partons. The medium-induced gluon radiation responsible for describing 2 → 3 processes uses the same higher twist formulation as that employed in MATTER.

In both modules, recoil partons that were created from elastic scatterings with the medium are transferred back to the framework and interacts with the QGP using the appropriate modules. Once partons reach low virtuality and low energy, they are handed back to PYTHIA for hadronization.

The modular nature of JETSCAPE allows for different energy loss formalism to work together. What energy loss mechanism is used to simulate the evolution of the parton depends on the magnitude of its energy and virtuality, which can change in both directions. This is a key feature of the JETSCAPE framework. 

\section{Results}

The charmed cross section in p+p collisions is explored in Ref.~\cite{gojko2020}. The agreement between PYTHIA and CMS data is not as good as when using PYTHIA combined with MATTER for final state radiation. With this improved baseline calculation, we further tune the separation scale $Q_0$ from 2~GeV to $\sqrt{2}$~GeV to calculate $D_0$ meson $R_{AA}$ and $v_2\{2\}$ with the aforementioned hydrodynamic background.

\begin{figure}[htbp]
	\centering
	\includegraphics[width=1\textwidth]{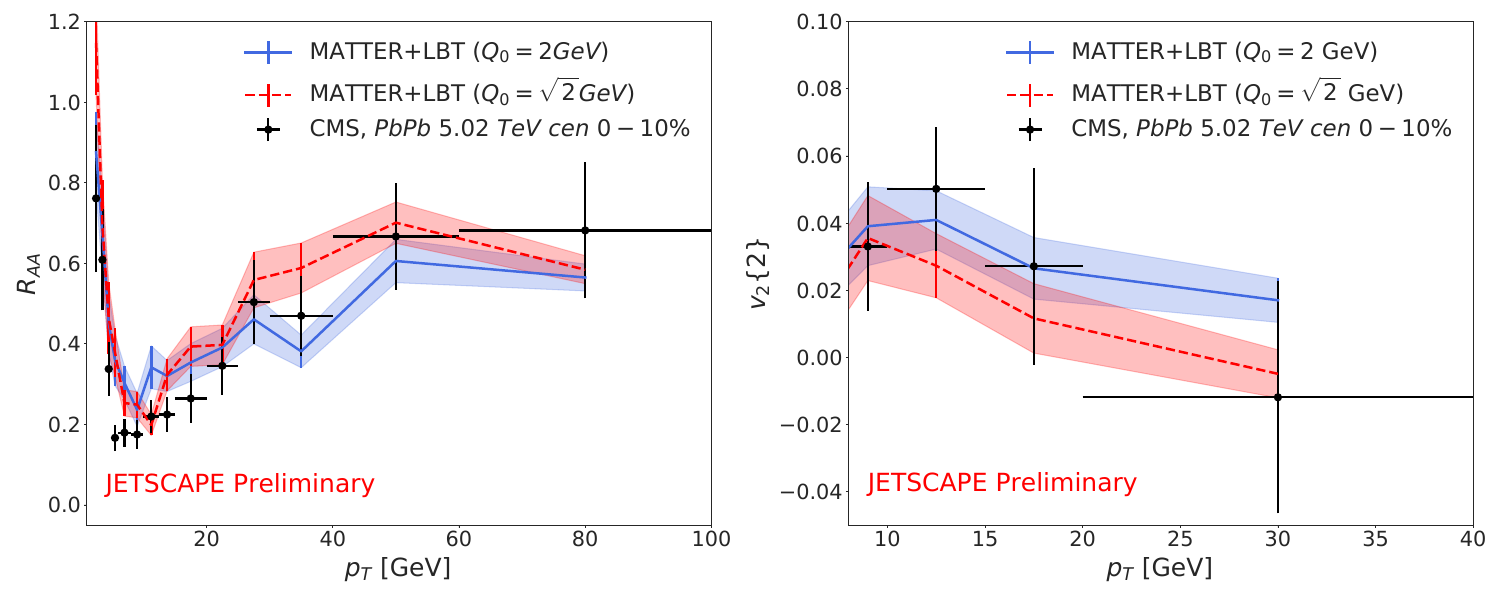}
	\caption{Comparison of $D^0$ meson nuclear modification factor $R_{AA}$ and azimuthal anisotropy $v_2\{2\}$ using the multi-stage energy loss approach (MATTER + LBT) with CMS data for central Pb + Pb collisions at $5.02$ TeV. (Left) $D^0$ meson $R_{AA}$.  (Right) $D^0$ meson $v_2\{2\}$. Computed with two switching virtuality $Q_0= 2  $ GeV and $Q_0= \sqrt{2}$ GeV.}
	\label{Dmesonresult}
\end{figure}

As we can see from Fig.~\ref{Dmesonresult}, both the $D^0$ meson $R_{AA}$ and $v_2\{2\}$ are described well using the same parameter set ($\alpha_s=0.25$ and $Q_0=2$~GeV) that is used to describe many light flavor and jet observables \cite{amit2019, yasuki2018}. A smaller $Q_0=\sqrt{2}$~GeV can still fit the data pretty well, as our predictions should be insensitive to the switching virtuality within a reasonable range. One have to note that $v_2\{2\}$ data below 10~GeV is not shown here as previous study \cite{cao2015} has shown the lack of a recombination mechanism for hadronization can lead to a significant suppression in heavy quark flow at low $p_T$. We will adopt a more detailed implementation of hadronization in the future.

\section{Conclusion}

We have presented an event-by-event calculation of $D^0$ meson $R_{AA}$ and $v_2$ using a multi-stage transport model within the JETSCAPE framework. Good agreement with CMS data is obtained, partly stemming from the improvement of the p+p baseline when using PYTHIA+MATTER in generating the parton shower in the vacuum. We have found that a single parameter set can describe many light flavor, heavy flavor and jet observables simultaneously. We will continue to explore more physics to further improve our description in the hard sector of heavy ion collisions. Future work for heavy flavor observables will include a better description of heavy quark hadronization in the low $p_T$ range and also study for bottom flavor observables. 

\section*{Acknowledgement}

This work has been supported by the U.S DOE DE-FG02-05ER41367 and NSF ACI-1550300.

\end{document}